\documentclass[5pt]{article}
\usepackage{amsmath}
\usepackage[latin1]{inputenc}
\usepackage{amsmath,amsthm,amssymb}
\usepackage{graphicx}
\usepackage{color}
\usepackage{multirow}
\usepackage{rotating}
\usepackage{subeqnarray}
\usepackage{cases}

\newcommand{\tabincell}[2]{\begin{tabular}{@{}#1@{}}#2\end{tabular}}
\renewcommand{\theequation}{\thesection.\arabic{equation}}

\title{ Asymmetric Single Magnitude Four Error Correcting Codes}
\author{{Derong Xie  \qquad\quad    Jinquan Luo}\footnote{ The authors are with school of mathematics and
statistics \& Hubei Key Laboratory of Mathematical Sciences,
Central China Normal University
Wuhan 430079, China.
E-mail: luojinquan@mail.ccnu.edu.cn(J.Luo); derongxie@yahoo.com.
}}
\date{}
\begin{document}
\baselineskip15pt \maketitle
\renewcommand{\theequation}{\arabic{section}.\arabic{equation}}
\catcode`@=11 \@addtoreset{equation}{section} \catcode`@=12

$\mathbf{Abstract-}$Limited magnitude asymmetric error model is well suited for flash memory. In this paper, we consider the construction of asymmetric codes correcting single error over $\mathbb{Z}_{2^{k}r}$ and which are based on so called $B_{1}[4](2^{k}r)$ set. In fact, we reduce the construction of a maximal size $B_{1}[4](2^{k}r)$ set for $k\geq3$ to the construction of a maximal size $B_{1}[4](2^{k-3}r)$ set. Finally, we give a explicit formula of a maximal size $B_{1}[4](4r)$ set and some lower bounds of a maximal size $B_{1}[4](2r)$ set. By computer searching up to $q\leq106$, we conjecture that those lower bounds are tight.

\emph{$\mathbf{Index \ Terms-}$}Asymmetric error,\  single error,\ flash memories,\ limited magnitude error.

\section{Introduction}

Flash memory is a kind of non-volatile memory which has higher transfer speed, longer life span and less sensitive of vibration than hard disks. But the material of flash memory is expensive and has fixed blocks, which makes it necessary to increase the density of flash memory.
At the same time, it faces many challenges such as how to implement codes correcting asymmetric errors into the flash memories. In \cite{AAK}, the asymmetric channel with limited magnitude errors was introduced and the further results were given in \cite{AAKT1,AAKT2}. An error model with asymmetric errors of limited magnitude is a good model for some multilevel flash memories. In the asymmetric error model, a symbol $a$ over an alphabet
$$\mathbb{Z}_{q}=\{0,1,\cdots,q-1\}$$
may be modified during transmission into $b$, where $b\geq a$, and the probability that $a$ is changed to $b$ is considered to be the same for all $b>a$. For some applications, the error magnitude $b-a$ is not likely to exceed a certain level $\lambda.$  In general, the errors are mostly asymmetric and some classes of construction of asystematic codes correcting such errors were studied in \cite{CSB,KBN,KLNY}.
Also, several constructions of systematic codes correcting single errors are given in \cite{KBN}
and the symmetric case is closely related to equi-difference conflict-avoiding codes see e.g., \cite{XL,ZZG}. In addition, splitter sets can be seen as codes correcting single limited magnitude errors in flash memories see e.g., \cite{BE,KLNY,KLY,M,YKB,ZG,ZZG}.

On the other hand, construction of codes correcting $t$ errors can be transformed to $B_{t}[\lambda](q)$ sets and the construction of a maximal size $B_{1}[3](2^{k}r)$ set, $B_{1}[3](3^{k}r)$ set and $B_{1}[4](3^{k}r)$ set can be found in \cite{KLNY}. In this paper, we consider the construction of a maximal size $B_{1}[4](2^{k}r)$ set. In Section \uppercase\expandafter{\romannumeral2}, we briefly introduce $B_{1}[\lambda](q)$ set and linear codes over the ring $\mathbb{Z}_{q}$. Indeed, we recall some basic results on $B_{1}[\lambda](q)$ sets. In Section \uppercase\expandafter{\romannumeral3}, we reduce the construction of a maximal size $B_{1}[4](2^{k}r)$ set for $k\geq3$ to the construction of a maximal size $B_{1}[4](2^{k-3}r)$ set. In Section \uppercase\expandafter{\romannumeral4}, we give an exact formula for calculating a maximal size $B_{1}[4](4r)$ set. In section \uppercase\expandafter{\romannumeral5}, we consider maximal size $B_{1}[4](2r)$ set. Finally, we give a short summary of this paper in Section \uppercase\expandafter{\romannumeral6}.

\section{Preliminaries}

The following result is almost identical to the introduction given in \cite{KBN,KLNY,KLY}. But we include it here for the completeness of this paper.

If $H$ is an $h\times m$ matrix over $\mathbb{Z}_{q}$, the corresponding code of length $m$ with parity check matrix $H$, is
$$\mathcal{C}_{H}=\{\mathbf{x}\in\mathbb{Z}_{q}^{m} \ | \ \mathbf{x}H^{t}=\mathbf{0}\}$$
where $H^{t}$ denotes the transposed of $H$.

Let $\mathcal{E}\subset\mathbb{Z}_{q}^{m}$ be the set of error patterns that we want to
correct and consider single errors of magnitude at most $\lambda$. If $\mathbf{x}\in \mathcal{C}_{H}$ is a sent codeword and $\mathbf{e}\in\mathcal{E}$ is an error introduced during transmission, then the received $m$-tuple is $\mathbf{y}=\mathbf{x}+\mathbf{e}$. Therefore
$$\mathbf{y}H^{t}=\mathbf{x}H^{t}+\mathbf{e}H^{t}=\mathbf{e}H^{t}.$$
As usual, we call $\mathbf{e}H^{t}$ the syndrome of $\mathbf{e}$. Let
$$\mathcal{S}_{H,\mathcal{E}}=\{\mathbf{e}H^{t} \ | \mathbf{e}\in\mathcal{E}\}$$
be the set of syndromes. We require these to be all distinct, i.e., $|\mathcal{S}_{H,\mathcal{E}}|=|\mathcal{E}|$. When this is the case, the code is able to correct all error patterns in $\mathcal{E}$. Moreover,
$$\bigcup_{\mathbf{x}\in\mathcal{C}_{H}}\{\mathbf{x}+\mathbf{e} \ | \ \mathbf{e}\in\mathcal{E}\}$$
is a disjoint union, and so we get the Hamming type bound
$$|\mathcal{C}_{H}|\cdot|\mathcal{E}|=|\mathcal{C}_{H}|\cdot|\mathcal{S}_{H,\mathcal{E}}|\leq q^{m}.$$

For ordinary linear codes (for $q$ a prime power), when $h=1$, $\mathcal{C}_{H}$ is an $[m.m-1]$ code of minimum Hamming distance two that
an not correct any errors (without limitation on the magnitude). When we consider errors of limited magnitude, the situation may be quite different, and it is a nontrivial task to find good $H$. Therefore, we consider $h=1$, that is $H=(b_{0},b_{1},\cdots,b_{m-1})$, and the error patterns we consider are
$\mathcal{E}_{\lambda,m}$, the set of sequences $(e_{0},e_{1},\cdots,e_{m-1})\in [0,\lambda]^{m}$ of Hamming weight at most 1. We see that
$$|\mathcal{E}_{\lambda,m}|=m\lambda+1.$$

Permuting the elements of $(b_{0},b_{1},\cdots,b_{m-1})$, we get another code with the same error correcting capability. Therefore, from now on we consider sets
$$B=\{b_{0},b_{1},\cdots,b_{m-1}\}$$
of distinct positive integers such that the corresponding syndromes
$$\mathcal{S}=\left\{\sum_{j=0}^{m-1}e_{j}b_{j} \ (\mbox{mod}\ q)\ |\ (e_{0},e_{1},\cdots,e_{m-1})\in\mathcal{E}_{\lambda,m}\right\}$$
are distinct. This is called a $B_{1}[\lambda](q)$ set, see \cite{KBN}. The corresponding code we denote by $\mathcal{C}_{B}$, that is
$$\mathcal{C}_{B}=\left\{(x_{0},x_{1},\cdots,x_{m-1})\in\mathbb{Z}_{q}^{m} \ \left|\ \sum_{i=0}^{m-1}x_{i}b_{i}\equiv0 \ (\mbox{mod}\ q)\right.\right\}.$$

For any positive integer $l$ coprime to $d$, let $\mbox{ord}_{d}(l)$ be the order of $l$ in $\mathbb{Z}_{d}^{*}$, that is,
$$\mbox{ord}_{d}(l)=\min\{n>0 \ | \ l^{n}\equiv1 \ (\mbox{mod}\ d)\}.$$
For $a\in\mathbb{Z}_{d}^{*}$ and $d$ a divisor of $q$, we let $\beta=aq/d.$

For $\gcd(al,d)=1$, define the cyclotomic set
$$\sigma_{l}(\beta)=\{l^{i}\beta \ (\mbox{mod}\ q) \ |\ i\geq 0 \}.$$
Then $|\sigma_{l}(\beta)|=\mbox{ord}_{d}(l).$

$Lemma \ 1:$ (\cite{KLNY}, Lemma 5) \ a) For $d=p_{1}^{e_{1}}p_{2}^{e_{2}}\cdots p_{s}^{e_{s}}$ with $p_{i}$ distinct primes not diving $l$, we have
$$\mbox{ord}_{d}=\mbox{lcm}\left(\mbox{ord}_{p_{1}^{e_{1}}}(l),\mbox{ord}_{p_{2}^{e_{2}}}(l),\cdots,\mbox{ord}_{p_{s}^{e_{s}}}(l)\right).$$
b) If $p$ is a prime not dividing $l$ and $l^{\mbox{ord}_{p}(l)}-1=p^{\mu_{p}}a$, where $\gcd(a,p)=1$, then
$$
\mbox{ord}_{p^{k}}(l)=
\left\{
\begin{array}{ll}
\mbox{ord}_{p}(l) & if \ k\leq\mu_{p},\\
p^{k-\mu_{p}}\mbox{ord}_{p}(l) & if \ k>\mu_{p},  \ p=2 \ \mbox{and} \ l\equiv1(\mbox{mod}\ 4),\\
p^{k-\mu_{p}}\mbox{ord}_{p}(l) & if \ k>\mu_{p} \ \mbox{and} \ p>2,\\
2 & if  \ p=2,\ l\equiv3(\mbox{mod}\ 4) \ \mbox{and} \ k=2,3,\\
2^{k-2} & if \ p=2,\ l\equiv3(\mbox{mod}\ 4) \ \mbox{and} \ k>3.
\end{array}
\right.
$$

\section{Maximal size $B_{1}[4](2^{k}r)$ set}

For $q=2^{k}r$ with $\gcd(r,6)=1$, we will introduce a result reducing the construction of a maximal size $B_{1}[4](2^{k}r)$ set for $k\geq3$ to the construction of a maximal size $B_{1}[4](2^{k-3}r)$ set in this section. Define $M_{4}(q)$ to be the maximal size of a $B_{1}[4](q)$ set.

For a positive integer $q=2^{k}r$ with $\gcd(r,6)=1$ and $d\mid r$, let
$$V_{d}=\{ar/d \ (\mbox{mod} \ q)\ | \ a  \in \mathbb{Z}_{2^{k}d}, \ \gcd(a,d)=1\} \ \mbox{and}\ U_{i}=\{x\in V_{d} \mid \gcd(x,2^{k})=2^{i}\}.$$
We have
\begin{equation}\label{def VdUi}
\mathbb{Z}_{2^{k}r}=\bigcup_{d\mid r}V_{d}\ \mbox{and}\ V_{d}=\bigcup_{0\leq i\leq k}U_{i}.
\end{equation}

Let
\begin{equation}\label{def M'}
M_{4}^{'}(2^{k}d)=\mbox{max}\left\{\left|\{B\cap V_{d} \ |\ B\in B_{1}[4](2^{k}r)\}\right|\right\}.
\end{equation}
Then
$$M_{4}(2^{k}r)=\sum_{d\mid r}M_{4}^{'}(2^{k}d).$$

 If $k\geq3$, we consider the following disjoint decomposition
 $$Z_{q}\backslash\{0\}=N_{0}\cup N_{1}\cup N_{2}\cup N_{3},$$
 where
$$N_{i}=\{a \ \mbox{mod}\ q \mid 1\leq a\leq 2^{k-i}r, 2\nmid a\} \ \mbox{for} \ 0\leq i \leq2$$
and
$$N_{3}=\{8a \ \mbox{mod}\ q \mid 1\leq a< 2^{k-3}r\}.$$

$Theorem \ 1:$ If $\gcd(r,6)=1$ and $k\geq 3$, then we have
$$M_{4}(2^{k}r)=M_{4}(2^{k-3}r)+2^{k-3}r.$$
$Proof:$ For any $d\mid r$, write $n_{k}=\mbox{ord}_{2^{k}d}(3)$ for short if no confusion occurs. Then $n_{k-1}$ is even and $n_{k-1}=n_{k-2}$ or $n_{k-1}=2n_{k-2}$.
For any $a$, let $\alpha=ar/d$ and the value of $\alpha$ varies with $a$. If $\gcd(a,2^{k}d)=1,$ we have
$$\left| \sigma_{3}(\alpha)\right|=\mbox{ord}_{2^{k}d}(3)=n_{k}.$$
Let $\Gamma_{2^{k}d}$ be a set of coset representatives of the group generated by $3$ in $\mathbb{Z}_{2^{k}d}^{*}$. It suffices to consider the following cases.

(i) If $n_{k-2}$ is odd, then $n_{k-1}=2n_{k-2}$ and $k=3$. Hence,
$$n_{k}= \mbox{lcm} \left( \mbox{ord}_{2^{3}}(3), \mbox{ord}_{d}(3) \right)=\mbox{lcm}\left( \mbox{ord}_{2^{2}}(3), \mbox{ord}_{d}(3) \right)=n_{k-1}.$$
We can choose
$$T_{d}=\bigcup_{a\in \Gamma_{2^{k-2}d}}\left\{ 3^{2i}\alpha \ (\mbox{mod}\ 2^{k}r) \mid 0\leq i \leq n_{k-2}-1 \right\}.$$

(ii) If $n_{k-2}$ is even and $n_{k-1}=n_{k-2}$, then $1+2^{k-2}d$ is not in the cyclic  group generated by 3 in $\mathbb{Z}_{2^{k-1}d}^{*}$. We can choose
   $$T_{d}= \bigcup_{a\in \Gamma_{2^{k-3}d}}\left(\left\{ 3^{2i}\alpha \ (\mbox{mod}\ 2^{k}r) \mid 0\leq i \leq \frac{n_{k-2}}{2}-1 \right\}\cup \left\{ 3^{2i+1}(\alpha+2^{k-2}r) \ (\mbox{mod}\ 2^{k}r) \mid 0\leq i \leq \frac{n_{k-2}}{2}-1 \right\}\right).$$

(iii) It is easy to see that $n_{k-2}$ is even and $n_{k-1}=2n_{k-2}$ is equivalent to $k\geq5$ and $v_{2}\left( \mbox{ord}_{d}(3)\right)\leq k-4$,  which implies
$$n_{k}=2n_{k-1}.$$

In this case, then we can choose
$$T_{d}= \bigcup_{a\in \Gamma_{2^{k-2}d}}\left(\left\{ 3^{2i}\alpha \ (\mbox{mod}\ 2^{k}r) \mid 0\leq i \leq \frac{n_{k-2}}{2}-1 \right\}\cup \left\{ 3^{2i+1}\alpha \ (\mbox{mod}\ 2^{k}r) \mid \frac{n_{k-2}}{2}\leq i \leq n_{k-2}-1 \right\}\right).$$

In total, we choose $S_{0}=\bigcup_{d\mid r}T_{d}$. Let $S_{1}$ be a $B_{1}[4](2^{k-3}r)$ set. Define
$$S=S_{0}\cup S^{'} \ \mbox{with} \ S^{'}=\{8c \ (\mbox{mod}\ 2^{k}r) \mid c\in S_{1}\}.$$
Obviously $S^{'}$ is a  $B_{1}[4](2^{k}r)$ set. Indeed, it is easy to verify that there do not exist distinct elements $x,y\in S_{0}$ such that $2x\equiv2y \ (\mbox{mod}\ 2^{k}r)$, $3x\equiv3y \ (\mbox{mod}\ 2^{k}r)$ or $4x\equiv4y \ (\mbox{mod}\ 2^{k}r)$. Hence $2S$, $3S$ and $4S$ both have the same size as $S$. Note that $S^{'}$ is a $B_{1}[4](2^{k}r)$ set contained in $N_{3}$. We see that
\begin{itemize}
  \item  $S\cap2S=\emptyset$ since $S_{0}\subset N_{0}$, $2S_{0}\subset N_{1}$.
  \item  $S\cap3S=\emptyset$ since $S_{0}\subset N_{0}$, $3S_{0}\subset N_{0}$ and $S_{0}\cap 3S_{0}=\emptyset$.
  \item  $S\cap4S=\emptyset$ since $S_{0}\subset N_{0}$, $4S_{0}\subset N_{2}$.
  \item  $2S\cap3S=\emptyset$ since $2S_{0}\subset N_{1}$, $3S_{0}\subset N_{0}$.
  \item  $2S\cap4S=\emptyset$ since $2S_{0}\subset N_{1}$, $4S_{0}\subset N_{2}$.
  \item  $3S\cap4S=\emptyset$ since $3S_{0}\subset N_{0}$, $4S_{0}\subset N_{2}$.
\end{itemize}

Then $S$ is a $B_{1}[4](2^{k}r)$ set of size $M_{4}(2^{k-3}r)+2^{k-3}r$.

On the other hand, firstly we note that at least one of $a,2a,3a,4a$ belongs to $N_{2}$ for any $a\in N_{0}\cup N_{1}\cup N_{2}$. Therefore, in $N_{0}\cup N_{1}\cup N_{2}$, at most $|N_{2}|=2^{k-3}r$ elements can be chosen in a $B_{1}[4](2^{k}r)$ set.
Also, neither of $a,2a,3a,4a$ belongs to $N_{0}\cup N_{1}\cup N_{2}$ for any $a\in N_{3}$. Since
$$(S_{0}\cup 2S_{0}\cup 3S_{0}\cup 4S_{0})\subset (N_{0}\cup N_{1}\cup N_{2}),$$
then the set $S$ is a maximal size $B_{1}[4](2^{k}r)$ set.\hfill$\blacksquare$\\

$Example \ 1:$
\begin{itemize}
\item For $q=40$, we have
$$M_{4}(40)=M_{4}(5)+5=6.$$
The construction of maximal set in the proof of Theorem 1 is presented as follows. Firstly we have $k=3$ and $r=5$. Hence, $d=1$ or $d=5$.\\
$-$If $d=1$, then $n_{2}=2$, $n_{1}=1$, and Case (i) applies. We can choose $\Gamma_{2}=\{1\}$ and so $T_{1}=\{5\}$.\\
$-$If $d=5$, then $n_{2}=2$, $n_{1}=4$, and Case (ii) applies. Choose $\Gamma_{5}=\{1\}$. Then
\begin{eqnarray*}
T_{5}&=&\left\{ 3^{2i} \ (\mbox{mod}\ 40) \mid 0\leq i \leq 1 \right\}\cup \left\{ 3^{2i+1}\cdot11 \ (\mbox{mod}\ 40) \mid 0\leq i \leq 1 \right\}\\
&=& \{1,9,33,17\}.
\end{eqnarray*}
We can choose $\{1\}$ as a perfect $B_{1}[4](5)$ set. Then we obtain a maximal size $B_{1}[4](40)$ set
$$T_{1}\cup T_{5}\cup 8\cdot\{1\}=\{1,5,8,9,17,33\}.$$
\item For $q=160$, note that $\{1,9,13,17\}$ is a maximal size $B_{1}[4](20)$ set (see Example 2). Therefore
$$M_{4}(160)=20+M_{4}(20)=24.$$
The construction of maximal set in the proof of Theorem 1 is depicted as follows. Firstly we have $k=5$ and $r=5$. Hence, $d=1$ or $d=5$.\\
$-$If $d=1$, then $n_{4}=4$, $n_{3}=2$, and Case (iii) applies. We can choose $\Gamma_{8}=\{1,5\}$ and so
\begin{eqnarray*}
T_{1}&=&\left\{ 1\cdot5 \ (\mbox{mod}\ 160), \ 3^{3}\cdot5 \ (\mbox{mod}\ 160)\right\}\cup \left\{ 1\cdot25 \ (\mbox{mod}\ 160), \ 3^{3}\cdot25 \ (\mbox{mod}\ 160) \right\}\\
&=&\{5,135,25,35\}.
\end{eqnarray*}
$-$If $d=5$, then $n_{4}=4$, $n_{3}=4$, and Case (ii) applies. Choose $\Gamma_{20}=\{1,7,11,13\}$. Then
\begin{eqnarray*}
T_{5}&=&\left\{ 3^{2i} \ (\mbox{mod}\ 160) \mid 0\leq i \leq 1 \right\}\cup \left\{ 3^{2i+1}\cdot41 \ (\mbox{mod}\ 160) \mid 0\leq i \leq 1 \right\}\\
&&\cup\left\{ 3^{2i}\cdot7 \ (\mbox{mod}\ 160) \mid 0\leq i \leq 1 \right\}\cup \left\{ 3^{2i+1}\cdot47 \ (\mbox{mod}\ 160) \mid 0\leq i \leq 1 \right\}\\
&&\cup\left\{ 3^{2i}\cdot11 \ (\mbox{mod}\ 160) \mid 0\leq i \leq 1 \right\}\cup \left\{ 3^{2i+1}\cdot51 \ (\mbox{mod}\ 160) \mid 0\leq i \leq 1 \right\}\\
&&\cup\left\{ 3^{2i}\cdot13 \ (\mbox{mod}\ 160) \mid 0\leq i \leq 1 \right\}\cup \left\{ 3^{2i+1}\cdot53 \ (\mbox{mod}\ 160) \mid 0\leq i \leq 1 \right\}\\
&=& \{1, 9, 123, 147, 7, 63, 141, 149, 11, 99, 153, 97, 13, 117, 159, 151\}.
\end{eqnarray*}
Hence,
\begin{eqnarray*}
T_{1}\cup T_{5}\cup 8\cdot\{1,9,13,17\}&=&\{1, 5, 7, 8, 9, 11, 13, 25, 35, 63, 72, 97, 99, 104,\\
 &&117, 123, 135, 136, 141, 147, 149, 151, 153, 159\}
\end{eqnarray*}
is a maximal size $B_{1}[4](160)$ set.
\end{itemize}

\section{Construction of maximal size $B_{1}[4](4r)$ sets}

In this section, we give an explicit construction of maximal size $B_{1}[4](4r)$ sets. We consider the following disjoint decomposition
 $$\mathbb{Z}_{4r}\backslash\{0\}=N_{0}\cup N_{1}\cup N_{2}\cup N_{3}$$
 where
\begin{equation}
\begin{aligned}
 & N_{0}=\{a \ \mbox{mod}\ 4r \mid 1\leq a\leq 4r, 2\nmid a\}\\
 & N_{1}=\{2a \ \mbox{mod}\ 4r \mid 1\leq a\leq 2r, 2\nmid a\}\\
 & N_{2}=\{4a \ \mbox{mod}\ 4r \mid 1\leq a< r\}.\nonumber
 \end{aligned}
\end{equation}

$Theorem \ 2:$ If $\gcd(r,6)=1$, then
$$M_{4}(4r)=r-1.$$

$Proof:$  1) If $r=1$, then $M_{4}=0.$

 2) If $r>1$ with $\gcd(r,6)=1$, for any $d\mid r$ and $d\geq5$ , let $\Gamma_{4d}$ be a set of coset representatives of the group generated by $3$ in $\mathbb{Z}_{4d}^{*}$. It suffices to consider the following cases.

 (i) If $n_{1}=\mbox{ord}_{2d}(3)$ is odd, then $n_{1}=n_{0}$ and $n_{2}=2n_{1}$. We can choose
$$T_{d}=\bigcup_{a\in \Gamma_{2d}}\left\{ 3^{2i}\alpha \ (\mbox{mod}\ 4r) \mid 0\leq i < n_{1} \right\}.$$
For distinct elements $3^{2i}\alpha \ (\mbox{mod}\ 4r), \ 3^{2i'}\alpha \ (\mbox{mod}\ 4r)\in T_{d}$, since $n_{0}$ is odd, then $3^{2(i-i^{'})}\not\equiv1\ (\mbox{mod}\ d)$ and so $4\cdot3^{2i}\alpha \ (\mbox{mod}\ 4r)\not\equiv4\cdot3^{2i}\alpha \ (\mbox{mod}\ 4r).$

 (ii) If $n_{1}=\mbox{ord}_{2d}(3)$ is even, then $n_{2}=n_{1}=n_{0}$ and $1+2d$ is not in the cyclic group generated by 3 in $\mathbb{Z}_{4d}^{*}$. We have
 $$\alpha(1+2d)\equiv\alpha+2r \ (\mbox{mod}\ 4r),$$
 $$2\cdot3^{i}\alpha(\alpha+2r)\equiv2\cdot3^{i}\alpha \ (\mbox{mod}\ 4r),$$
  $$4\cdot3^{i}\alpha(\alpha+2r)\equiv4\cdot3^{i}\alpha \ (\mbox{mod}\ 4r).$$
  We can choose
  $$T_{d}= \bigcup_{a\in \Gamma_{2d}}\left(\left\{ 3^{2i}\alpha \ (\mbox{mod}\ 4r) \mid 0\leq i < \frac{n_{1}}{2} \right\}\cup \left\{ 3^{2i+1}(\alpha+2r) \ (\mbox{mod}\ 4r) \mid 0\leq i < \frac{n_{1}}{2} \right\}\right).$$

  In total, we choose $S=\bigcup_{d>1,d\mid r}T_{d}$.
  It is easy to verify that there do not exist distinct elements $x,y\in S_{0}$ such that $2x\equiv2y \ (\mbox{mod}\ 2^{k}r)$, $3x\equiv3y \ (\mbox{mod}\ 2^{k}r)$ or $4x\equiv4y \ (\mbox{mod}\ 2^{k}r)$. Hence $2S$, $3S$ and $4S$ both have the same size as $S$. Obviously, $S\cap 3S=\emptyset$. Since $S\subset N_{0}$, $2S\subset N_{1}$,  $3S\subset N_{0}$  and $4S\subset N_{2}$, then $S\cap 2S=\emptyset$, $S\cap 4S=\emptyset$, $2S\cap 3S=\emptyset$, $2S\cap 4S=\emptyset$ and $3S\cap 4S=\emptyset$.

  Therefore, $S$ is a $B_{1}[4](4r)$ set of size $r-1$.

  On the other hand,
  $$M_{4}(4r)\leq \left\lfloor\frac{4r-1}{4}\right\rfloor=r-1.$$
Hence, the set $S$ is a maximal size $B_{1}[4](4r)$ set.\hfill$\blacksquare$\\

$Remark\ 1:$ Combining Theorems 1 and 2, if $k\equiv2 \ (\mbox{mod}\ 3)$, then
$$M_{4}(2^{k}r)=\frac{1}{7}(2^{k}r+3r-7).$$

$Example \ 2:$
\begin{itemize}
\item For $q=20$, we have
$$M_{4}(20)=5-1=4.$$
The construction of maximal set in the proof of Theorem 2 is presented as follows. Firstly we have $r=5$. Hence, $d=1$ or $d=5$.
If $d=5$, then $n_{1}=4$ is even and Case (ii) applies. We can choose $\Gamma_{10}=\{1\}$ and so
\begin{eqnarray*}
T_{5}&=&\left\{ 3^{2i} \ (\mbox{mod}\ 20) \mid 0\leq i \leq 1 \right\}\cup \left\{ 3^{2i+1}\cdot21 \ (\mbox{mod}\ 20) \mid 0\leq i \leq 1 \right\}\\
&=& \{1,9,13,17\}
\end{eqnarray*}
is a  maximal size $B_{1}[4](20)$ set.
\item For $q=44$, we have
$$M_{4}(20)=11-1=10.$$
The construction of maximal set in the proof of Theorem 2 is depicted as follows. Firstly we have $r=11$. Hence, $d=1$ or $d=11$.
If $d=11$, then $n_{1}=5$ is odd and Case (i) applies. We can choose $\Gamma_{22}=\{1,7\}$ and so
\begin{eqnarray*}
T_{11}&=&\left\{ 3^{2i} \ (\mbox{mod}\ 44) \mid 0\leq i \leq 4 \right\}\cup \left\{ 3^{2i}\cdot7 \ (\mbox{mod}\ 44) \mid 0\leq i \leq 4 \right\}\\
&=& \{1, 9, 37, 25, 5, 7, 19, 39, 43, 35\}
\end{eqnarray*}
is a  maximal size $B_{1}[4](20)$ set.
\end{itemize}

\section{On lower bounds of $M_{4}^{'}(2d)$}

Let $\langle3\rangle_{d}$ be the group generated by 3 in $\mathbb{Z}_{2^{k}d}^{*}$. Recall $V_{d}$ in \ref{def VdUi} and $M_{4}^{'}(2d)$ in \ref{def M'}. We give a formula or lower bound for $M_{4}^{'}(2d)$ whether $2\in\langle3\rangle_{d}$ or not.
Firstly, $M_{4}^{'}(2)=0.$ For $q=2r$, we note that $|U_{0}|=|U_{1}|=\varphi(d)$ where $\varphi$ is the $Euler^{,}s \  totient$ function.
Let $\theta_{2}: \mathbb{Z}_{2r}\rightarrow\mathbb{Z}_{2r}$  defined by $\theta_{2}(x)=2x \ (\mbox{mod}\ 2r).$

$Lemma \ 2:$ By restricting the domain of the mapping $\theta_{2}$ to $U_{i}$ of $\mathbb{Z}_{2r}$, we have

(1) $\theta_{2}(U_{0})=U_{1}$, which is a bijection.

(2) $\theta_{2}(U_{1})=U_{1}$, which is a bijection.

$Proof:$  For $x\in \mathbb{Z}_{2r}$, let $t=x+r \ (\mbox{mod}\ 2r)$, then $x\neq t$ and $\theta_{2}(x)=\theta_{2}(t)=2x  \ (\mbox{mod}\ 2r)$.

 For any $2a\in U_{1}$, we have two cases to consider: if $a$ is odd, then $a\in U_{0}, t\in U_{1}$ and $\theta_{2}(2a)=\theta_{2}(t)=2a(\mbox{mod}\ 2r)$; if $a$ is even, then $t\in U_{0}, 2a(\mbox{mod}\ 2r)\in U_{1}$ and $\theta_{2}(2a)=\theta_{2}(t)=2a (\mbox{mod}\ 2r)$. Since $|U_{0}|=|U_{1}|$, then $\theta_{2}|_{U_{0}}$ and $\theta_{2}|_{U_{1}}$  are both bijections.\hfill$\blacksquare$\\
 
\subsection{On lower bounds of $M_{4}^{'}(2d)$ with $2\in\langle3\rangle_{d}$ and $d\geq5$}

For brevity, we let
\begin{itemize}
\item $n=\mbox{ord}_{d}(3)$;
\item $2\equiv3^{s} \ (\mbox{mod}\ d)$ with $s\in[1,n]$;
\item $m=\min\{s,n-s\}$;
\item $n=2k^{'}m+r^{'}$ with $0\leq r^{'}<2m$;
\end{itemize}

$Theorem \ 3:$ If $n$ is even and $s$ is odd, then
$$M_{4}{'}(2d)=\frac{1}{2}\varphi(d).$$

$Proof :$ If $n$ is even and $s$ is odd, we can choose
$$T_{d}=\bigcup_{a\in \Gamma_{2d}}\left\{ 3^{2i}\alpha \ (\mbox{mod}\ 2r) \mid 0\leq i < \frac{n}{2} \right\}.$$

 We note that $T_{d}\subset U_{0}$. For distinct elements $x,y\in T_{d}$, we have $2x\not\equiv2y \ (\mbox{mod}\ 2r)$ and $4x\not\equiv4y \ (\mbox{mod}\ 2r)$ by Lemma 2. Clearly, $3x\not\equiv3y \ (\mbox{mod}\ 2r)$. Hence $2T_{d}, 3T_{d}$ and $4T_{d}$ both have the same size as $T_{d}$. Obviously, $T_{d}\cap 3T_{d}=\emptyset$. Checking binary parity we can get $T_{d}\cap 2T_{d}=\emptyset, \ T_{d}\cap 4T_{d}=\emptyset, \ 2T_{d}\cap 3T_{d}=\emptyset$ and $3T_{d}\cap 4T_{d}=\emptyset$. Since $s\in [1,n]$ is odd and $i,j\in[0,n/2-1]$, then $3^{2i}\not\equiv2\cdot3^{2j} \ (\mbox{mod}\ d)$ which implies that $2T_{d}\cap 4T_{d}=\emptyset$. Hence $S$ is a $B_{1}[4](2r)$ set. Therefore,
$$M_{4}^{'}(2d)\geq|T_{d}|=\frac{1}{2}\varphi(d).$$

On the other hand, it is clear that
$$M_{4}^{'}(2d)\leq \frac{|V_{d}|}{4}=\frac{1}{2}\varphi(d)$$
and so $M_{4}^{'}(2d)=\frac{1}{2}\varphi(d)$.\hfill$\blacksquare$\\

$Example \ 3:$  For $q=2\cdot5\cdot19=190$, we have $r=95$. Hence, $d=1$, $d=5$, $d=19$ or $95$.\\
$-$If $d=5$, then $s=3$ and $n=4$. We have $|M_{4}^{'}(10)|=\varphi(5)/2=2$. Choose $\Gamma_{10}=\{1\}$ and so
$$T_{5}=\left\{ 3^{2i}\cdot19 \ (\mbox{mod}\ 190) \mid 0\leq i < 2 \right\}
=\{19,171\}.$$
$-$If $d=19$, then $s=7$ and $n=18$. We have $|M_{4}^{'}(38)|=\varphi(19)/2=9$. Choose $\Gamma_{38}=\{1\}$. Then
$$T_{19}=\left\{ 3^{2i}\cdot5 \ (\mbox{mod}\ 190) \mid 0\leq i < 9 \right\}
=\{5, 45, 25, 35, 125, 175, 55, 115, 85\}.$$
$-$If $d=95$, then $s=7$ and $n=36$. We have $|M_{4}^{'}(190)|=\varphi(95)/2=36$. Choose $\Gamma_{190}=\{1,7\}$. Then
\begin{eqnarray*}
T_{95}&=&\left\{ 3^{2i} \ (\mbox{mod}\ 160) \mid 0\leq i < 18 \right\}\cup\left\{ 3^{2i}\cdot7 \ (\mbox{mod}\ 160) \mid 0\leq i < 18 \right\}\\
&=&\left\{1, 9, 81, 159, 101, 149, 11, 99, 131, 39, 161, 119, 121, 139, 111, 49, 61, 169\right.\\
&& \left.7, 63, 187, 163, 137, 93, 77, 123, 157, 83, 177, 73, 87, 23, 17, 153, 47, 43\right\}.
\end{eqnarray*}
Hence, $M_{4}(190)=\sum_{d\mid 95}M_{4}^{'}(2d)=47$ and
\begin{eqnarray*}
T_{5}\cup T_{19}\cup T_{95}&=&\{1, 5, 7, 9, 11, 17, 19, 23, 25, 35, 39, 43, 45, 47, 49, 55, 61, 63, 73,\\
&&77, 81, 83, 85, 87, 93, 99, 101, 111, 115, 119, 121, 123, 125,131,\\
&& 137, 139, 149, 153, 157, 159, 161, 163, 169, 171, 175, 177, 187\}
\end{eqnarray*}
is a maximal size $B_{1}[4](190)$ set.\\

$Theorem \ 4:$ If both $n$ and $s$ are even , then\\
(1) in the case $m=2$, $$M_{4}{'}(2d)=\frac{\varphi(d)}{n}\cdot\left\lfloor\frac{n}{3}\right\rfloor;$$
(2) in the case $m>2$,
$$M_{4}{'}(2d)\geq
\left\{
\begin{array}{ll}
\frac{(m-1)\varphi(d)}{2m} & \mbox{if} \ r^{'}=0,\\
\frac{(k^{'}m-k^{'}+1)\varphi(d)}{n} & \mbox{if} \ r^{'}=2,\\
\frac{m\varphi(d)}{n} & \mbox{if} \ 2<r^{'}\leq m\ \mbox{and} \ k^{'}=1,\\
\frac{(k^{'}m-k^{'}+2)\varphi(d)}{n} & \mbox{if} \ 2<r^{'}\leq m \ \mbox{and} \ k^{'}\geq2,\\
\frac{(k^{'}m+r^{'}-m-k^{'})\varphi(d)}{n} & \mbox{if} \ r^{'}>m .
\end{array}
\right.$$

$Proof :$ (1) For $m=2$, we note that $4\cdot3^{i}\alpha\equiv 2\cdot3^{i+2}\alpha \ (\mbox{mod}\ 2r)$ or $2\cdot3^{i}\alpha\equiv4\cdot3^{i+2}\alpha  \ (\mbox{mod}\ 2r)$. Hence, if $3^{i}\alpha$ is chosen in a $B_{1}[4](2r)$ set, then
$3^{i+1}\alpha$ and $3^{i+2}\alpha$ can not be chosen. Therefore,
$$M_{4}^{'}(2d)\leq\frac{\varphi(d)}{n}\cdot\left\lfloor\frac{n}{3}\right\rfloor.$$

By Lemma 2, we know that $\theta_{2}|_{U_{0}}$ and $\theta_{2}|_{U_{1}}$  are both bijections. Hence, for any $z_{1}, \ z_{2}\in[1,4]$, $i,i^{'}<3\cdot\left\lfloor\frac{n}{3}\right\rfloor$ and $3\mid(i-i^{'})$, we have $z_{1}\cdot3^{3i}\alpha\equiv z_{2}\cdot3^{3i^{'}}\alpha \ (\mbox{mod}\ 2r)$ if and only if $z_{1}=z_{2}$ and $i=i^{'}$.
 Hence, we can choose
$$T_{d}=\bigcup_{a\in \Gamma_{2d}}\left\{ 3^{3i}\alpha \ (\mbox{mod}\ 2r) \mid 0\leq i < \left\lfloor\frac{n}{3}\right\rfloor \right\}.$$
Since $0\not\in T_{d}\cup2T_{d}\cup3T_{d}\cup4T_{d}$
and $|T_{d}|=\frac{\varphi(d)}{n}\cdot\left\lfloor\frac{n}{3}\right\rfloor$,
then $T_{d}$ is the maximal size of the elements of $B_{1}[4](2r)$ set in $V_{d}$.

(2) For $m>2:$

(i) If $r^{'}=0$, we can choose
$$T_{d}=\bigcup_{j=0}^{k^{'}-1}\bigcup_{a\in \Gamma_{2d}}\left(\left\{ 3^{2i}\alpha \ (\mbox{mod}\ 2r) \mid jm\leq i <\frac{m}{2}+jm \right\}\cup\left\{ 3^{2i+1}\alpha \ (\mbox{mod}\ 2r) \mid \frac{m}{2}+jm\leq i < m+jm-1 \right\}\right).$$

For distinct elements $x,y\in T_{d}$, it is easy to verify that $2x\not\equiv2y \ (\mbox{mod}\ 2r)$, $3x\not\equiv3y \ (\mbox{mod}\ 2r)$ and $4x\not\equiv4y \ (\mbox{mod}\ 2r)$. Hence $2T_{d}, 3T_{d}$ and $4T_{d}$ both have the same size as $T_{d}$. Obviously, $T_{d}\cap 3T_{d}=\emptyset$. Checking binary parity we can get $T_{d}\cap 2T_{d}=\emptyset, \ T_{d}\cap 4T_{d}=\emptyset, \ 2T_{d}\cap 3T_{d}=\emptyset$ and $3T_{d}\cap 4T_{d}=\emptyset$. For any
$$x\in\left\{ 3^{2i}\alpha \ (\mbox{mod}\ 2r) \mid jm\leq i <\frac{m}{2}+jm \right\}\cup\left\{ 3^{2i+1}\alpha \ (\mbox{mod}\ 2r) \mid \frac{m}{2}+jm\leq i < m+jm-1 \right\},$$
$$y\in\left\{ 3^{2i}\alpha \ (\mbox{mod}\ 2r) \mid j^{'}m\leq i <\frac{m}{2}+j^{'}m \right\}\cup\left\{ 3^{2i+1}\alpha \ (\mbox{mod}\ 2r) \mid \frac{m}{2}+j^{'}m\leq i < m+j^{'}m-1 \right\},$$
we obtain $x=3^{e_{1}}\alpha \ (\mbox{mod}\ 2r)$ and $y=3^{e_{2}}\alpha \ (\mbox{mod}\ 2r)$.
For $j=j^{'}$, if $e_{1}, e_{2}$ have the same binary parity
then $0<|ind_{3_{\alpha}}(x)-ind_{3_{\alpha}}(y)\pm m|<n$ and so $2x\not\equiv4y \ (\mbox{mod}\ 2r)$; otherwise, we also have $2x\not\equiv4y \ (\mbox{mod}\ 2r)$ by checking binary parity of $e_{1}, e_{2}$.

By Lemma 2, we know that $\theta_{2}$ is a bijection from $U_{1}$ to $U_{1}$.
If $|j-j^{'}|>1$, then $2x\not\equiv4y \ (\mbox{mod}\ 2r)$
since $2\cdot3^{t}\equiv 4\cdot3^{t+m} \ (\mbox{mod}\ 2r)$ or $4\cdot3^{t}\equiv2\cdot3^{t+m}  \ (\mbox{mod}\ 2r)$.

Similarly, $2x\not\equiv4y \ (\mbox{mod}\ 2r)$  holds for $|j-j^{'}|=1$. Therefore,
$$M_{4}^{'}(2d)\geq|T_{d}|=\frac{(m-1)\varphi(d)}{2m}.$$

(ii)If $r^{'}=2$, we can choose
\begin{eqnarray*}
&&T_{d}=\bigcup_{a\in \Gamma_{2d}}\Bigg(\bigcup_{j=0}^{k^{'}-1}\bigg(\left\{ 3^{2i}\alpha \ (\mbox{mod}\ 2r) \mid jm\leq i <\frac{m}{2}+jm \right\}\\
&\cup&\left\{ 3^{2i+1}\alpha \ (\mbox{mod}\ 2r) \mid \frac{m}{2}+jm\leq i < m+jm-1 \right\}\bigg)\cup\{3^{n-1}\alpha \ (\mbox{mod}\ 2r)\}\Bigg)
\end{eqnarray*}
and so
$$M_{4}{'}(2d)\geq\frac{(k^{'}m-k^{'}+1)\varphi(d)}{n}.$$

(iii)The case $2<r^{'}\leq m$
\begin{itemize}
\item For $k^{'}=1$, we can choose
$$
T_{d}=\bigcup_{a\in \Gamma_{2d}}\left(\{ 3^{2i}\alpha \ (\mbox{mod}\ 2r) \mid0\leq i < \frac{m}{2}\}
\cup \{ 3^{2i+1}\alpha \ (\mbox{mod}\ 2r) \mid \frac{m}{2}\leq i < m \}\right)
$$
and so
$$M_{4}{'}(2d)\geq\frac{m\varphi(d)}{n}.$$
\item For $k^{'}\geq2$,
we can choose
\begin{eqnarray*}
T_{d}&=&\bigcup_{a\in \Gamma_{2d}}\Bigg(\bigcup_{j=0}^{k^{'}-2}\bigg(\left\{ 3^{2i}\alpha \ (\mbox{mod}\ 2r) \mid jm\leq i <\frac{m}{2}+jm \right\}\\
&\cup& \left\{ 3^{2i+1}\alpha \ (\mbox{mod}\ 2r) \mid \frac{m}{2}+jm\leq i < m+jm-1 \right\}\bigg)\\
&\cup& \left\{ 3^{2i}\alpha \ (\mbox{mod}\ 2r) \mid k^{'}m-m+1\leq i <k^{'}m-\frac{m}{2}+1 \right\}\\
&\cup& \left\{ 3^{2i+1}\alpha \ (\mbox{mod}\ 2r) \mid k^{'}m-\frac{m}{2}+1\leq i \leq k^{'}m\right\}\\
&\cup& \left\{3^{2k^{'}m-2m-1}\alpha \ (\mbox{mod}\ 2r)\right\}\Bigg)
\end{eqnarray*}
\end{itemize}
and so
$$M_{4}{'}(2d)\geq\frac{(k^{'}m-k^{'}+2)\varphi(d)}{n}.$$

If $r^{'}>m $, we can choose
\begin{eqnarray*}
T_{d}&=&\bigcup_{a\in \Gamma_{d}}\Bigg(\bigcup_{j=0}^{k^{'}-2}\bigg(\left\{ 3^{2i}\alpha \ (\mbox{mod}\ 2r) \mid jm\leq i <\frac{m}{2}+jm \right\}\\
&\cup& \left\{ 3^{2i+1}\alpha \ (\mbox{mod}\ 2r) \mid \frac{m}{2}+jm\leq i < m+jm-1 \right\}\bigg)\\
&\cup& \left\{ 3^{2i}\alpha \ (\mbox{mod}\ 2r) \mid k^{'}m-m\leq i <k^{'}m+\frac{r^{'}-3m}{2} \right\}\\
&\cup& \left\{ 3^{2i+1}\alpha \ (\mbox{mod}\ 2r) \mid k^{'}m-\frac{m+2}{2}\leq i < k^{'}m+\frac{r^{'}-2m-2}{2}\right\}\\
&\cup& \left\{ 3^{2i}\alpha \ (\mbox{mod}\ 2r) \mid k^{'}m+\frac{r^{'}-2m}{2}\leq i <k^{'}m+\frac{r^{'}-m}{2} \right\}\\
&\cup& \left\{ 3^{2i+1}\alpha \ (\mbox{mod}\ 2r) \mid k^{'}m+\frac{r^{'}-m}{2}\leq i < k^{'}m+\frac{r^{'}}{2}\right\}\Bigg)
\end{eqnarray*}
and so
$$M_{4}{'}(2d)\geq\frac{(k^{'}m+r^{'}-m-k^{'})\varphi(d)}{n}.$$\hfill$\blacksquare$\\

$Lemma \ 3:$ If $n$ is odd, then $m\neq1,2.$

$Proof:$  We assume that $m=1$ which implies $2\equiv3 \ (\mbox{mod}\ d)$ or $2\equiv3^{n-1} \ (\mbox{mod}\ d)$. Then $d=5$ and $\mbox{ord}_{5}(3)=4$
which contradicts to that $n$ is odd. Similarly, if $m=2$, then $d=7,17$ and $\mbox{ord}_{7}(3)=6,\ \mbox{ord}_{17}(3)=16$.
\hfill$\blacksquare$\\

$Theorem \ 5:$ The following holds for $n$ being odd.\\
(1) If $m$ is odd, then
$$M_{4}^{'}(2d)\geq\frac{(n-m)\varphi(d)}{2n}.$$
(2) If $m$ is even, then
$$M_{4}{'}(2d)\geq
\left\{
\begin{array}{ll}
\frac{(n+r^{'}-m-2k^{'})\varphi(d)}{2n} & \mbox{if} \ r^{'}\leq m,\\
\frac{(2k^{'}m+m-2k^{'}+1)\varphi(d)}{2n} & \mbox{if} \ r^{'}>m.
\end{array}
\right.$$

$Proof :$
(1) If $m$ is odd, similar to Theorem 3, we can choose
$$T_{d}=\bigcup_{a\in \Gamma_{2d}}\left\{ 3^{2i}\alpha \ (\mbox{mod}\ 2r) \mid 0\leq i < \frac{n-m}{2} \right\}$$
and so
$$M_{4}^{'}(2d)\geq|T_{d}|=\frac{(n-m)\varphi(d)}{2n}.$$

(2) If $m$ is even, similar to Theorem 4,
for $r^{'}\leq m$, we can choose
\begin{eqnarray*}
T_{d}&=&\bigcup_{a\in \Gamma_{2d}}\Bigg(\bigcup_{j=0}^{k^{'}-2}\bigg(\left\{ 3^{2i}\alpha \ (\mbox{mod}\ 2r) \mid jm\leq i <\frac{m}{2}+jm \right\}\\
&\cup& \left\{ 3^{2i+1}\alpha \ (\mbox{mod}\ 2r) \mid \frac{m}{2}+jm\leq i < m+jm-1 \right\}\bigg)\\
&\cup& \left\{ 3^{2i}\alpha \ (\mbox{mod}\ 2r) \mid k^{'}m-m\leq i <k^{'}m-\frac{m}{2} \right\}\\
&\cup& \left\{ 3^{2i+1}\alpha \ (\mbox{mod}\ 2r) \mid k^{'}m-\frac{m}{2}\leq i < \frac{n-m-1}{2}\right\}\\
&\cup& \left\{3^{2i}\alpha \ (\mbox{mod}\ 2r) \mid k^{'}m\leq i <\frac{n-1}{2}\right\}\Bigg)
\end{eqnarray*}
and so
$$M_{4}{'}(2d)\geq|T_{d}|=\frac{(n+r^{'}-m-2k^{'})\varphi(d)}{2n}.$$

If $r^{'}>m $, we can choose
\begin{eqnarray*}
T_{d}&=&\bigcup_{a\in \Gamma_{d}}\Bigg(\bigcup_{j=0}^{k^{'}-2}\bigg(\left\{ 3^{2i}\alpha \ (\mbox{mod}\ 2r) \mid jm\leq i <\frac{m}{2}+jm \right\}\\
&\cup& \left\{ 3^{2i+1}\alpha \ (\mbox{mod}\ 2r) \mid \frac{m}{2}+jm\leq i < m+jm-1 \right\}\bigg)\\
&\cup& \left\{ 3^{2i}\alpha \ (\mbox{mod}\ 2r) \mid k^{'}m-m\leq i <k^{'}m-\frac{m}{2} \right\}\\
&\cup& \left\{ 3^{2i+1}\alpha \ (\mbox{mod}\ 2r) \mid k^{'}m-\frac{m}{2}\leq i < k^{'}m\right\}\\
&\cup& \left\{ 3^{2i}\alpha \ (\mbox{mod}\ 2r) \mid \frac{n-m-1}{2}\leq i \leq \frac{n-3}{2}\right\}\Bigg)
\end{eqnarray*}
and so
$$M_{4}{'}(2d)\geq|T_{d}|=\frac{(2k^{'}m+m-2k^{'}+1)\varphi(d)}{2n}.$$\hfill$\blacksquare$\\
If $r$ is a prime, then $M_{4}(2r)=M_{4}^{'}(2r)$. For prime $r<100$, we give all the examples by Theorems 3,4 and 5 in Table 1.
\begin{table}[!h]
\caption{$M_{4}(2p)$ with $2\in\langle3\rangle_{p}$}
\centering
\begin{tabular}{|c|c|c|c|c|c|c|c|}
\hline
\hline
 $p$ & $n$ & $m$ & $k$ & $r^{'}$ & $M_{4}(2p)$ & $T_{d}$ & Thm. \\
 \hline
5 & 4 & 1 &  &  & 2 & $\{1,9\}$ & 3 \\
7 & 6 & 2 & 1 & 2 & 2 & $\{1,13\}$ & 4-(1) \\
17 & 16 & 2 & 4 & 0 & 5 & $\{1, 15, 21, 27, 31\}$ & 4-(1) \\
19 & 18 & 7 &  &  & 9 & $\{1, 5, 7, 9, 11, 17, 23, 25, 35\}$ & 3 \\
23 & 11 & 4 & 1 & 3 & $\geq8$ & $\{1, 5, 7, 9, 13, 19, 29, 45\}$ & 5-(2) \\
29 & 28 & 11 &  &  & 14 & $\{1, 5, 7, 9, 13, 23, 25, 33, 35, 45, 49, 51, 53, 57\}$ & 3 \\
31 & 30 & 6 & 2 & 6 & $\geq12$ & $\{1, 9, 11, 13, 15, 17, 19, 29, 35, 37, 41, 59\}$ &4 (2) \\
43 & 42 & 15 &  &  & 21 & \tabincell{c}{$\left\{1, 9, 11, 13, 15, 17, 21, 23, 25, 31, 35,\right.$ \\
$\left. 41, 47, 49, 53, 57, 59, 67, 79, 81, 83\right\}$} & 3 \\
47 & 23 & 6 & 1 & 11 & $\geq18$ & \tabincell{c}{$\left\{1, 5, 7, 9, 19, 25, 29, 31, 35, 37,\right.$ \\
$\left. 45, 51, 53, 67, 77, 79, 81, 91\right\}$} & 5-(2) \\
53 & 52 & 3 &  &  & 26 & \tabincell{c}{$\left\{1, 7, 9, 11, 13, 15, 17, 25, 29, 37, 43, 47, 49, 57, 59,\right.$ \\
$\left. 63, 69, 77, 81, 89, 91, 93, 95, 97, 99, 105\right\}$} & 3 \\
71 & 35 & 11 & 1 & 13 & $\geq24$ & \tabincell{c}{$\left\{1, 7, 9, 11, 17, 19, 23, 29, 43, 61, 63, 65, 75, 77,\right.$\\
$\left. 81, 99, 103, 113, 119, 123, 125, 131, 133, 141\right\}$} & 5-(1) \\
79 & 78 & 4 & 9 & 6 & $\geq29$ & \tabincell{c}{$\left\{1, 7, 9, 13, 17, 19, 23, 31, 35, 41, 45, 61, 65, 67, 83, 85, 95,\right.$\\ $\left.99, 101, 103, 107, 109, 115, 121, 129, 131, 143, 147, 155\right\}$} & 4-(2) \\
89 & 88 & 16 & 2 & 24 & $\geq38$ & \tabincell{c}{$\left\{1, 5, 7, 9, 13, 17, 33, 41, 43, 55, 57, 63, 67, 69, 71, 75,\right.$\\
 77, 81, 87, 93, 95, 105,  109, 111, 113, 117, 125,\\
 $\left. 127, 131, 135, 139, 143, 147, 151, 153, 157, 159, 163\right\}$} & 4-(2)\\
 97 & 48 & 5 &  &  & $48$ & \tabincell{c}{$\left\{1, 5, 9, 13, 17, 19, 21, 23, 29, 33, 35, 41, 43, 45, 47,\right.$\\
  61, 67, 73, 75, 77, 81, 83, 91, 93, 101, 103, 111,\\
 113, 117, 119, 121, 127, 133, 147, 149, 151, 153,\\
 $\left. 159, 161, 165, 171, 173, 175, 177, 181, 185, 189, 193\right\}$} & 3\\
\hline
\end{tabular}
\end{table}

\subsection{On lower bounds of $M_{4}^{'}(2d)$ with $2\not\in\langle3\rangle_{d}$ and $d\geq5$}

Let $\Lambda_{d}$ be a set of coset representatives of $\langle2,3\rangle_{d}$ in $\mathbb{Z}_{d}^{*}$ with $1\in\Lambda_{d}$. Let $\Gamma_{2,3}$ be a set of coset representatives of the group generated by $3$ in $\langle2,3\rangle_{d}$ and so we can choose $\Gamma_{2,3}=\{1,2,\cdots,2^{t-1}\}$ where $t=\frac{|\langle2,3\rangle_{d}|}{|\langle3\rangle_{d}|}$. Since $|\langle3\rangle_{2d}|=|\langle3\rangle_{d}|$, then the natural map
\begin{align}
& \mathbb{Z}_{2d}^{*}\longrightarrow\mathbb{Z}_{d}^{*}\nonumber\\
& \quad x\longmapsto x \ (\mbox{mod} \ d)\nonumber
\end{align}
induces a group isomorphism $$\phi: \ \mathbb{Z}_{2d}^{*}/\langle3\rangle_{2d}\longrightarrow\mathbb{Z}_{d}^{*}/\langle3\rangle_{d}$$ and $\phi(d+2)=2$. Denote by $b=d-2$. Then
$$\bigcup_{a\in\Lambda_{d}}a\cdot\{1,b,\cdots,b^{t-1}\}$$
is a set of coset representatives of $\langle3\rangle_{2d}$ in $\mathbb{Z}_{2d}^{*}$. For $2\not\in\langle3\rangle_{d}$, there exists a unique integer $s\in[1,n-1]$ such that $4\cdot b^{t-1}\cdot3^{s}\equiv2 \ (\mbox{mod}\ 2d)$, i.e., $b^{t}\cdot3^{s}\equiv1 \ (\mbox{mod}\ d)$.

$Theorem \ 6:$ If both $n$ and $t+s$ are even, then
$$M_{4}{'}(2d)=\frac{1}{2}\varphi(d).$$

$Proof :$
\begin{itemize}
\item If both $t$ and $s$ are even, we can choose
\begin{eqnarray*}
T_{d}=\bigcup_{a\in \Lambda_{d}}\bigcup_{j=0}^{\frac{t-2}{2}}\left(\left\{ 3^{2i}b^{2j}\alpha \ (\mbox{mod}\ 2r) \mid 0 \leq i <\frac{n}{2} \right\}\cup\left\{ 3^{2i+1}b^{2j+1}\alpha \ (\mbox{mod}\ 2r) \mid 0 \leq i <\frac{n}{2}\right\}\right).
\end{eqnarray*}
\item If both $t$ and $s$ are odd, we can choose
\begin{eqnarray*}
T_{d}&=&\bigcup_{a\in \Lambda_{d}}\Bigg(\bigg(\bigcup_{j=0}^{\frac{t-1}{2}}\left\{ 3^{2i}b^{2j}\alpha \ (\mbox{mod}\ 2r) \mid 0 \leq i <\frac{n}{2} \right\}\bigg)\\
&&\cup\bigg(\bigcup_{j=0}^{\frac{t-3}{2}}\left\{ 3^{2i+1}b^{2j+1}\alpha \ (\mbox{mod}\ 2r) \mid 0 \leq i <\frac{n}{2}\right\}\bigg)\Bigg).
\end{eqnarray*}
For $i,i^{'}\in[0,n-1]$ and $j,j^{'}\in[0,t-1]$, if $(i,j)\neq(i^{'},j^{'})$, then
$$3^{i}2^{j}\not\equiv3^{i^{'}}2^{j^{'}}\ (\mbox{mod}\ d).$$
Therefore, for distinct elements $x,y\in T_{d}$, it is easy to verify that $2x\not\equiv2y \ (\mbox{mod}\ 2r)$, $3x\not\equiv3y \ (\mbox{mod}\ 2r)$ and $4x\not\equiv4y \ (\mbox{mod}\ 2r)$. Hence all of $2T_{d}, 3T_{d}$ and $4T_{d}$ have the same size as $T_{d}$. Obviously, $T_{d}\cap 3T_{d}=\emptyset$. Checking binary parity we can get $T_{d}\cap 2T_{d}=\emptyset, \ T_{d}\cap 4T_{d}=\emptyset, \ 2T_{d}\cap 3T_{d}=\emptyset$ and $3T_{d}\cap 4T_{d}=\emptyset$.
Indeed, if both $t$ and $s$ are even, then
$$\left\{ 2\cdot3^{2i}\alpha \ (\mbox{mod}\ 2r) \mid 0 \leq i <\frac{n}{2} \right\}
\cap\left\{4\cdot3^{2i+1}b^{t-1}\alpha \ (\mbox{mod}\ 2r) \mid 0 \leq i <\frac{n}{2} \right\}=\emptyset.$$
If both $t$ and $s$ are odd, then
$$\left\{ 2\cdot3^{2i}\alpha \ (\mbox{mod}\ 2r) \mid 0 \leq i <\frac{n}{2} \right\}
\cap\left\{4\cdot3^{2i}b^{t-1}\alpha \ (\mbox{mod}\ 2r) \mid 0 \leq i <\frac{n}{2} \right\}=\emptyset.$$
Hence, in combination with the proof of the Theorem 4, we have $2T_{d}\cap 4T_{d}=\emptyset.$
Therefore,
$$M_{4}^{'}(2d)\geq|T_{d}|=\frac{|\Lambda_{d}|\cdot n\cdot t}{2}=\frac{\varphi(d)}{2}.$$

On the other hand, it is clear that
$$M_{4}^{'}(2d)\leq \frac{|V_{d}|}{4}=\frac{\varphi(d)}{2}$$
and so
$$M_{4}^{'}(2d)=\frac{\varphi(d)}{2}.$$
\hfill$\blacksquare$\\

$Theorem \ 7:$ The following holds for $n$ being even.

(1) If $t$ is odd and $s$ is even, then
$$M_{4}{'}(2d)\geq
\left\{
\begin{array}{ll}
|\Lambda_{d}|\cdot\frac{(n-2)\cdot(t-1)+2s}{2} & \mbox{if} \ t-1<s<\frac{n}{2},\\
|\Lambda_{d}|\cdot\frac{ (n-2)\cdot t}{2} & \mbox{if} \ t<s=\frac{n}{2},\\
|\Lambda_{d}|\cdot\frac{ (n-2)\cdot (t-1)+2n-2s}{2} & \mbox{if} \ \frac{n}{2}<s<n-t+1,\\
|\Lambda_{d}|\cdot\frac{ n\cdot (t-1)}{2} & \mbox{otherwise}.
\end{array}
\right.$$

(2) If $t$ is even and $s$ is odd, then
$$M_{4}{'}(2d)\geq
\left\{
\begin{array}{ll}
|\Lambda_{d}|\cdot\frac{(n-2)\cdot(t-1)+2s+2}{2} & \mbox{if} \ t-2<s<\frac{n-2}{2},\\
|\Lambda_{d}|\cdot\frac{ (n-2)\cdot t}{2} & \mbox{if} \ t<\frac{n}{2} \ \mbox{and} \ s=\frac{n}{2} \ \mbox{or}\ \frac{n\pm2}{2},\\
|\Lambda_{d}|\cdot\frac{ (n-2)\cdot (t-1)+2n-2s+2}{2} & \mbox{if} \ \frac{n}{2}<s<n-t+2,\\
|\Lambda_{d}|\cdot\frac{ n\cdot (t-1)}{2} & \mbox{otherwise}.
\end{array}
\right.$$

$Proof :$ (1) The case $t$ is odd and $s$ is even.
\begin{itemize}
\item If $t-1<s<\frac{n}{2}$, we can choose
\begin{eqnarray*}
T_{d}&=&\bigcup_{a\in \Lambda_{d}}\Bigg(\bigcup_{j=0}^{\frac{t-3}{2}}\bigg(\left\{ 3^{2i}b^{2j}\alpha \ (\mbox{mod}\ 2r) \mid 0 \leq i <\frac{s}{2} \right\}\\
&&\cup\left\{ 3^{2i+1}b^{2j}\alpha \ (\mbox{mod}\ 2r) \mid \frac{s}{2}\leq i <\frac{n-2}{2}\right\}\\
&&\cup\left\{ 3^{2i+1}b^{2j+1}\alpha \ (\mbox{mod}\ 2r) \mid 0 \leq i <\frac{s}{2}\right\}\\
&&\cup\left\{ 3^{2i}b^{2j+1}\alpha \ (\mbox{mod}\ 2r) \mid \frac{s+2}{2}\leq i <\frac{n}{2}\right\}\bigg)\\
&&\cup\left\{ 3^{2i}b^{t-1}\alpha \ (\mbox{mod}\ 2r) \mid 0 \leq i <\frac{s}{2}\right\}\\
&&\cup\left\{ 3^{2i+1}b^{t-1}\alpha \ (\mbox{mod}\ 2r) \mid \frac{s}{2} \leq i <s\right\}\Bigg).
\end{eqnarray*}
Clearly, for any $a\in\Lambda_{d}$ and $j\in[0,\frac{t-3}{2}]$,
\begin{eqnarray*}
\!&&\!\!\!\!\bigg(\left\{ 4\cdot3^{2i}b^{2j}\alpha \ (\mbox{mod}\ 2r) \mid 0 \leq i <\frac{s}{2} \right\}\cup\left\{ 4\cdot3^{2i+1}b^{2j}\alpha \ (\mbox{mod}\ 2r) \mid \frac{s}{2}\leq i <\frac{n-2}{2}\right\}\bigg)\bigcap\\
\!&&\!\!\!\!\left(\left\{2\cdot 3^{2i+1}b^{2j+1}\alpha \ (\mbox{mod}\ 2r) \mid 0 \leq i <\frac{s}{2}\right\}\cup\left\{ 2\cdot3^{2i}b^{2j+1}\alpha \ (\mbox{mod}\ 2r) \mid \frac{s+2}{2}\leq i <\frac{n}{2}\right\}\right)=\emptyset.
\end{eqnarray*}
Indeed,
$$\left\{ 4\cdot3^{2i}b^{t-1}\alpha \ (\mbox{mod}\ 2r) \mid 0\leq i <\frac{s}{2}\right\}=\left\{ 2\cdot3^{2i}b\alpha \ (\mbox{mod}\ 2r) \mid \frac{n-s}{2} \leq i <\frac{n}{2}\right\}$$
and
$$\left\{ 4\cdot3^{2i+1}b^{t-1}\alpha \ (\mbox{mod}\ 2r) \mid \frac{s}{2} \leq i <s\right\}
=\left\{ 2\cdot3^{2i+1}b\alpha \ (\mbox{mod}\ 2r) \mid 0 \leq i <\frac{s}{2}\right\}.$$
Therefore,
$$M_{4}^{'}(2d)\geq|T_{d}|=|\Lambda_{d}|\cdot\frac{(n-2)\cdot(t-1)+2s}{2}.$$

\item If $t<s=\frac{n}{2}$, we can choose
\begin{eqnarray*}
T_{d}&=&\bigcup_{a\in \Lambda_{d}}\Bigg(\bigcup_{j=0}^{\frac{t-3}{2}}\bigg(\left\{ 3^{2i}b^{2j}\alpha \ (\mbox{mod}\ 2r) \mid 0 \leq i <\frac{s}{2} \right\}\\
&&\cup\left\{ 3^{2i+1}b^{2j}\alpha \ (\mbox{mod}\ 2r) \mid \frac{s}{2}\leq i <\frac{n-2}{2}\right\}\\
&&\cup\left\{ 3^{2i+1}b^{2j+1}\alpha \ (\mbox{mod}\ 2r) \mid 0 \leq i <\frac{s}{2}\right\}\\
&&\cup\left\{ 3^{2i}b^{2j+1}\alpha \ (\mbox{mod}\ 2r) \mid \frac{s+2}{2}\leq i <\frac{n}{2}\right\}\bigg)\\
&&\cup\left\{ 3^{2i}b^{t-1}\alpha \ (\mbox{mod}\ 2r) \mid 0 \leq i <\frac{s}{2}\right\}\\
&&\cup\left\{ 3^{2i+1}b^{t-1}\alpha \ (\mbox{mod}\ 2r) \mid \frac{s}{2} \leq i <s-1\right\}\Bigg)
\end{eqnarray*}
and $M_{4}^{'}(2d)\geq|T_{d}|=|\Lambda_{d}|\cdot\frac{ (n-2)\cdot t}{2}.$
\item If $\frac{n}{2}<s<n-t+1$, we can choose
\begin{eqnarray*}
T_{d}&=&\bigcup_{a\in \Lambda_{d}}\Bigg(\bigcup_{j=0}^{\frac{t-3}{2}}\bigg(\left\{ 3^{2i}b^{2j}\alpha \ (\mbox{mod}\ 2r) \mid 0 \leq i <\frac{n-s}{2} \right\}\\
&&\cup\left\{ 3^{2i+1}b^{2j}\alpha \ (\mbox{mod}\ 2r) \mid \frac{n-s}{2}\leq i <\frac{n-2}{2}\right\}\\
&&\cup\left\{ 3^{2i+1}b^{2j+1}\alpha \ (\mbox{mod}\ 2r) \mid 0 \leq i <\frac{n-s}{2}\right\}\\
&&\cup\left\{ 3^{2i}b^{2j+1}\alpha \ (\mbox{mod}\ 2r) \mid \frac{n-s+2}{2}\leq i <\frac{n}{2}\right\}\bigg)\\
&&\cup\left\{ 3^{2i}b^{t-1}\alpha \ (\mbox{mod}\ 2r) \mid 0 \leq i <\frac{n-s}{2}\right\}\\
&&\cup\left\{ 3^{2i+1}b^{t-1}\alpha \ (\mbox{mod}\ 2r) \mid \frac{s-2}{2} \leq i <\frac{n-2}{2}\right\}\Bigg)
\end{eqnarray*}
and $M_{4}^{'}(2d)\geq|T_{d}|=|\Lambda_{d}|\cdot\frac{ (n-2)\cdot (t-1)+2n-2s}{2}.$
\item Otherwise, we can choose
\begin{eqnarray*}
T_{d}=\bigcup_{a\in \Lambda_{d}}\bigcup_{j=0}^{\frac{t-3}{2}}\bigg(\left\{ 3^{2i}b^{2j}\alpha \ (\mbox{mod}\ 2r) \mid 0 \leq i <\frac{n}{2} \right\}\cup\left\{ 3^{2i+1}b^{2j+1}\alpha \ (\mbox{mod}\ 2r) \mid 0 \leq i <\frac{n}{2}\right\}\bigg)
\end{eqnarray*}
and $M_{4}^{'}(2d)\geq|T_{d}|=|\Lambda_{d}|\cdot\frac{ n\cdot (t-1)}{2}.$
\end{itemize}

(2) The case $t$ is even and $s$ is odd.
\begin{itemize}
\item If $t-2<s< \frac{n-2}{2}$, then we can choose
\begin{eqnarray*}
T_{d}&=&\bigcup_{a\in \Lambda_{d}}\Bigg(\Bigg(\bigcup_{j=0}^{\frac{t-2}{2}}\bigg(\left\{ 3^{2i+1}b^{2j}\alpha \ (\mbox{mod}\ 2r) \mid 0 \leq i <\frac{s+1}{2} \right\}\\
&&\cup\left\{ 3^{2i}b^{2j}\alpha \ (\mbox{mod}\ 2r) \mid \frac{s+3}{2}\leq i <\frac{n}{2}\right\}\bigg)\Bigg)\\
&&\cup\Bigg(\bigcup_{j=0}^{\frac{t-4}{2}}\bigg(\left\{ 3^{2i}b^{2j+1}\alpha \ (\mbox{mod}\ 2r) \mid 0 \leq i <\frac{s+1}{2} \right\}\\
&&\cup\left\{ 3^{2i+1}b^{2j+1}\alpha \ (\mbox{mod}\ 2r) \mid \frac{s+1}{2}\leq i <\frac{n-2}{2}\right\}\bigg)\Bigg)\\
&&\cup\left\{ 3^{2i}b^{t-1}\alpha \ (\mbox{mod}\ 2r) \mid 0 \leq i <\frac{s+1}{2}\right\}\\
&&\cup\left\{ 3^{2i+1}b^{t-1}\alpha \ (\mbox{mod}\ 2r) \mid \frac{s+1}{2} \leq i <s+1\right\}\Bigg)
\end{eqnarray*}
and $M_{4}^{'}(2d)\geq|T_{d}|=|\Lambda_{d}|\cdot\frac{ (n-2)\cdot (t-1)+2n-2s+2}{2}.$

\item If $t<\frac{n}{2}$ and $s=\frac{n-2}{2}$, we can choose
\begin{eqnarray*}
T_{d}&=&\bigcup_{a\in \Lambda_{d}}\Bigg(\Bigg(\bigcup_{j=0}^{\frac{t-2}{2}}\bigg(\left\{ 3^{2i+1}b^{2j}\alpha \ (\mbox{mod}\ 2r) \mid 0 \leq i <\frac{n}{4} \right\}\\
&&\cup\left\{ 3^{2i}b^{2j}\alpha \ (\mbox{mod}\ 2r) \mid \frac{n+4}{4}\leq i <\frac{n}{2}\right\}\bigg)\Bigg)\\
&&\cup\Bigg(\bigcup_{j=0}^{\frac{t-4}{2}}\bigg(\left\{ 3^{2i}b^{2j+1}\alpha \ (\mbox{mod}\ 2r) \mid 0 \leq i <\frac{n}{4} \right\}\\
&&\cup\left\{ 3^{2i+1}b^{2j+1}\alpha \ (\mbox{mod}\ 2r) \mid \frac{n}{4}\leq i <\frac{n-2}{2}\right\}\bigg)\Bigg)\\
&&\cup\left\{ 3^{2i}b^{t-1}\alpha \ (\mbox{mod}\ 2r) \mid 0 \leq i <\frac{n}{4}\right\}\\
&&\cup\left\{ 3^{2i+1}b^{t-1}\alpha \ (\mbox{mod}\ 2r) \mid \frac{n}{4} \leq i <\frac{n-2}{2}\right\}\Bigg)
\end{eqnarray*}
and $M_{4}^{'}(2d)\geq|T_{d}|=|\Lambda_{d}|\cdot\frac{ (n-2)\cdot t}{2}.$

\item If $t<s=\frac{n}{2}$, we can choose
\begin{eqnarray*}
T_{d}&=&\bigcup_{a\in \Lambda_{d}}\Bigg(\Bigg(\bigcup_{j=0}^{\frac{t-2}{2}}\bigg(\left\{ 3^{2i+1}b^{2j}\alpha \ (\mbox{mod}\ 2r) \mid 0 \leq i <\frac{s+1}{2} \right\}\\
&&\cup\left\{ 3^{2i}b^{2j}\alpha \ (\mbox{mod}\ 2r) \mid \frac{s+3}{2}\leq i <\frac{n}{2}\right\}\bigg)\Bigg)\\
&&\cup\Bigg(\bigcup_{j=0}^{\frac{t-4}{2}}\bigg(\left\{ 3^{2i}b^{2j+1}\alpha \ (\mbox{mod}\ 2r) \mid 0 \leq i <\frac{s+1}{2} \right\}\\
&&\cup\left\{ 3^{2i+1}b^{2j+1}\alpha \ (\mbox{mod}\ 2r) \mid \frac{s+1}{2}\leq i <\frac{n-2}{2}\right\}\bigg)\Bigg)\\
&&\cup\left\{ 3^{2i}b^{t-1}\alpha \ (\mbox{mod}\ 2r) \mid 0 \leq i <\frac{s+1}{2}\right\}\\
&&\cup\left\{ 3^{2i+1}b^{t-1}\alpha \ (\mbox{mod}\ 2r) \mid \frac{s+1}{2} \leq i <\frac{n-2}{2}\right\}\Bigg)
\end{eqnarray*}
and $M_{4}^{'}(2d)\geq|T_{d}|=|\Lambda_{d}|\cdot\frac{ (n-2)\cdot t}{2}.$

\item If $t<\frac{n}{2}$ and $s=\frac{n+2}{2}$, we can choose
\begin{eqnarray*}
T_{d}&=&\bigcup_{a\in \Lambda_{d}}\Bigg(\Bigg(\bigcup_{j=0}^{\frac{t-2}{2}}\bigg(\left\{ 3^{2i}b^{2j}\alpha \ (\mbox{mod}\ 2r) \mid 0 \leq i <\frac{n}{4} \right\}\\
&&\cup\left\{ 3^{2i+1}b^{2j}\alpha \ (\mbox{mod}\ 2r) \mid \frac{n+4}{4}\leq i <\frac{n}{2}\right\}\bigg)\Bigg)\\
&&\cup\Bigg(\bigcup_{j=0}^{\frac{t-4}{2}}\bigg(\left\{ 3^{2i+1}b^{2j+1}\alpha \ (\mbox{mod}\ 2r) \mid 0 \leq i <\frac{n}{4} \right\}\\
&&\cup\left\{ 3^{2i}b^{2j+1}\alpha \ (\mbox{mod}\ 2r) \mid \frac{n}{4}\leq i <\frac{n-2}{2}\right\}\bigg)\Bigg)\\
&&\cup\left\{ 3^{2i+1}b^{t-1}\alpha \ (\mbox{mod}\ 2r) \mid 0 \leq i <\frac{n}{4}\right\}\\
&&\cup\left\{ 3^{2i}b^{t-1}\alpha \ (\mbox{mod}\ 2r) \mid \frac{n}{4} \leq i <\frac{n-2}{2}\right\}\Bigg)
\end{eqnarray*}
and $M_{4}^{'}(2d)\geq|T_{d}|=|\Lambda_{d}|\cdot\frac{ (n-2)\cdot t}{2}.$

\item If $\frac{n+2}{2}<s<n-t+1$, we can choose
\begin{eqnarray*}
T_{d}&=&\bigcup_{a\in \Lambda_{d}}\Bigg(\Bigg(\bigcup_{j=0}^{\frac{t-2}{2}}\bigg(\left\{ 3^{2i}b^{2j}\alpha \ (\mbox{mod}\ 2r) \mid 0 \leq i <\frac{n-s+1}{2} \right\}\\
&&\cup\left\{ 3^{2i+1}b^{2j}\alpha \ (\mbox{mod}\ 2r) \mid \frac{n-s+1}{2}\leq i <\frac{n-2}{2}\right\}\bigg)\Bigg)\\
&&\cup\Bigg(\bigcup_{j=0}^{\frac{t-4}{2}}\bigg(\left\{ 3^{2i+1}b^{2j+1}\alpha \ (\mbox{mod}\ 2r) \mid 0 \leq i <\frac{n-s+1}{2} \right\}\\
&&\cup\left\{ 3^{2i}b^{2j}\alpha \ (\mbox{mod}\ 2r) \mid \frac{n-s+3}{2}\leq i <\frac{n}{2}\right\}\bigg)\Bigg)\\
&&\cup\left\{ 3^{2i+1}b^{t-1}\alpha \ (\mbox{mod}\ 2r) \mid 0 \leq i <\frac{n-s+1}{2}\right\}\\
&&\cup\left\{ 3^{2i}b^{t-1}\alpha \ (\mbox{mod}\ 2r) \mid \frac{s-1}{2} \leq i <\frac{n}{2}\right\}\Bigg)
\end{eqnarray*}
and $M_{4}^{'}(2d)\geq|T_{d}|=|\Lambda_{d}|\cdot\frac{ (n-2)\cdot (t-1)+2n-2s+2}{2} .$
\item Otherwise, we can choose
$$
T_{d}=\bigcup_{a\in \Lambda_{d}}\left(\bigg(\bigcup_{j=0}^{\frac{t-2}{2}}\left\{ 3^{2i}b^{2j}\alpha \ (\mbox{mod}\ 2r) \mid 0 \leq i <\frac{n}{2} \right\}\bigg)
\cup\bigg(\bigcup_{j=0}^{\frac{t-4}{2}}\left\{ 3^{2i+1}b^{2j+1}\alpha \ (\mbox{mod}\ 2r) \mid 0 \leq i <\frac{n}{2}\right\}\bigg)\right)
$$
and $M_{4}^{'}(2d)\geq|T_{d}|=|\Lambda_{d}|\cdot\frac{ n\cdot (t-1)}{2}.$
\end{itemize}
\end{itemize}
\hfill$\blacksquare$\\

$Theorem \ 8:$ The following holds for $n$ being odd.

(1) If $t$ is odd, then
$$M_{4}{'}(2d)\geq
\left\{
\begin{array}{ll}
|\Lambda_{d}|\cdot\frac{(n-1)\cdot t}{2} & \mbox{if} \ s=1,\\
|\Lambda_{d}|\cdot\frac{ (n-1)(t-1)+n-s+2}{2} & \mbox{if} \ s \ \mbox{is \ odd \ and}\ s>1,\\
|\Lambda_{d}|\cdot\frac{ (n-1)\cdot (t-1)+2}{2} & \mbox{if} \ s=0,\\
|\Lambda_{d}|\cdot\frac{ (n-1)(t-1)+s}{2} & \mbox{if} \ s \ \mbox{is \ even \ and}\ s>0.
\end{array}
\right.$$

(2) If $t$ is even, then
$$M_{4}{'}(2d)\geq
\left\{
\begin{array}{ll}
|\Lambda_{d}|\cdot\frac{(n-1)(t-1)+s+1}{2} & \mbox{if} \ s \ \mbox{is \ odd},\\
|\Lambda_{d}|\cdot\frac{ (n-1)\cdot t}{2} & \mbox{if} \ s=0,\\
|\Lambda_{d}|\cdot\frac{ (n-1)(t-1)+n-s+1}{2} & \mbox{if} \ s \ \mbox{is \ even \ and}\ s>0.
\end{array}
\right.$$

$Proof :$ It is similar to the proof of Theorems 6 and 7.

(1) The case $t$ is odd.
\begin{itemize}
\item If $s=1$, we can choose
\begin{eqnarray*}
T_{d}&=&\bigcup_{a\in \Lambda_{d}}\Bigg(\bigg(\bigcup_{j=0}^{\frac{t-1}{2}}\left\{ 3^{2i}b^{2j}\alpha \ (\mbox{mod}\ 2r) \mid 0 \leq i <\frac{n-1}{2} \right\}\bigg)\\
&&\cup\bigg(\bigcup_{j=0}^{\frac{t-3}{2}}\left\{ 3^{2i+1}b^{2j+1}\alpha \ (\mbox{mod}\ 2r) \mid 0 \leq i <\frac{n-1}{2}\right\}\bigg)\Bigg)
\end{eqnarray*}
and $M_{4}^{'}(2d)\geq|T_{d}|=|\Lambda_{d}|\cdot\frac{(n-1)\cdot t}{2}.$
\item If $s$ is odd and $s>1$, then
$$\left\{2\cdot3^{2i}\alpha \ (\mbox{mod}\ 2r) \mid 0\leq i < \frac{n-1}{2}\right\}\cap\left\{4\cdot3^{2i}b^{t-1}\alpha \ (\mbox{mod}\ 2r) \mid \frac{s-1}{2} \leq i \leq \frac{n-1}{2}\right\}=\emptyset.$$
We can choose
\begin{eqnarray*}
T_{d}&=&\bigcup_{a\in \Lambda_{d}}\Bigg(\bigcup_{j=0}^{\frac{t-3}{2}}\bigg(\left\{ 3^{2i}b^{2j}\alpha \ (\mbox{mod}\ 2r) \mid 0 \leq i <\frac{n-1}{2} \right\}\\
&&\cup\left\{ 3^{2i+1}b^{2j+1}\alpha \ (\mbox{mod}\ 2r) \mid 0 \leq i <\frac{n-1}{2}\right\}\bigg)\\
&&\cup\left\{ 3^{2i}b^{t-1}\alpha \ (\mbox{mod}\ 2r) \mid \frac{s-1}{2} \leq i \leq \frac{n-1}{2}\right\}\Bigg)
\end{eqnarray*}
and $M_{4}^{'}(2d)\geq|T_{d}|=|\Lambda_{d}|\cdot\frac{ (n-1)(t-1)+n-s+2}{2}.$
\item If $s=0$, we can choose
\begin{eqnarray*}
T_{d}&=&\bigcup_{a\in \Lambda_{d}}\bigcup_{j=0}^{\frac{t-3}{2}}\Bigg(\left\{ 3^{2i}b^{2j}\alpha \ (\mbox{mod}\ 2r) \mid 0 \leq i <\frac{n-1}{2} \right\}\\
&&\cup\left\{ 3^{2i+1}b^{2j+1}\alpha \ (\mbox{mod}\ 2r) \mid 0 \leq i <\frac{n-1}{2}\right\}\\
&&\cup\left\{ 3^{n-1}b^{t-1}\alpha \ (\mbox{mod}\ 2r)\right\}\Bigg)
\end{eqnarray*}
and $M_{4}^{'}(2d)\geq|T_{d}|=|\Lambda_{d}|\cdot\frac{ (n-1)\cdot (t-1)+2}{2}.$
\item If $s$ is even and $s>0$, then
$$\left\{2\cdot3^{2i}\alpha \ (\mbox{mod}\ 2r) \mid 0\leq i < \frac{n-1}{2}\right\}\cap\left\{4\cdot3^{2i}b^{t-1}\alpha \ (\mbox{mod}\ 2r) \mid 0 \leq i <\frac{s}{2}\right\}=\emptyset.$$
We can choose
\begin{eqnarray*}
T_{d}&=&\bigcup_{a\in \Lambda_{d}}\bigcup_{j=0}^{\frac{t-3}{2}}\Bigg(\left\{ 3^{2i}b^{2j}\alpha \ (\mbox{mod}\ 2r) \mid 0 \leq i <\frac{n-1}{2} \right\}\\
&&\cup\left\{ 3^{2i+1}b^{2j+1}\alpha \ (\mbox{mod}\ 2r) \mid 0 \leq i <\frac{n-1}{2}\right\}\\
&&\cup\left\{ 3^{2i}b^{t-1}\alpha \ (\mbox{mod}\ 2r) \mid 0 \leq i <\frac{s}{2}\right\}\Bigg)
\end{eqnarray*}
and $M_{4}^{'}(2d)\geq|T_{d}|=|\Lambda_{d}|\cdot\frac{ (n-1)(t-1)+s}{2}.$
\end{itemize}

(2) The case $t$ is even.
\begin{itemize}
\item If $s$ is odd, then
$$\left\{2\cdot3^{2i+1}\alpha \ (\mbox{mod}\ 2r) \mid 0\leq i < \frac{n-1}{2}\right\}\cap\left\{4\cdot3^{2i}b^{t-1}\alpha \ (\mbox{mod}\ 2r) \mid 0 \leq i \leq \frac{s-1}{2}\right\}=\emptyset.$$
We can choose
\begin{eqnarray*}
T_{d}&=&\bigcup_{a\in \Lambda_{d}}\Bigg(\bigg(\bigcup_{j=0}^{\frac{t-2}{2}}\left\{ 3^{2i+1}b^{2j}\alpha \ (\mbox{mod}\ 2r) \mid 0 \leq i <\frac{n-1}{2} \right\}\bigg)\\
&&\cup\bigg(\bigcup_{j=0}^{\frac{t-4}{2}}\left\{ 3^{2i}b^{2j+1}\alpha \ (\mbox{mod}\ 2r) \mid 0 \leq i <\frac{n-1}{2}\right\}\bigg)\\
&&\cup\left\{ 3^{2i}b^{t-1}\alpha \ (\mbox{mod}\ 2r) \mid 0 \leq i < \frac{s+1}{2}\right\}\Bigg)
\end{eqnarray*}
and $M_{4}^{'}(2d)\geq|T_{d}|=|\Lambda_{d}|\cdot\frac{(n-1)(t-1)+s+1}{2}.$
\item If $s=0$, we can choose
\begin{eqnarray*}
T_{d}&=&\bigcup_{a\in \Lambda_{d}}\Bigg(\bigcup_{j=0}^{\frac{t-2}{2}}\left\{ 3^{2i+1}b^{2j}\alpha \ (\mbox{mod}\ 2r) \mid 0 \leq i <\frac{n-1}{2} \right\}\\
&&\cup\left\{ 3^{2i}b^{2j+1}\alpha \ (\mbox{mod}\ 2r) \mid 0 \leq i <\frac{n-1}{2}\right\}\Bigg)
\end{eqnarray*}
and $M_{4}^{'}(2d)\geq|T_{d}|=|\Lambda_{d}|\cdot\frac{ (n-1)\cdot t}{2}.$
\item If $s$ is even and $s>0$, then
$$\left\{2\cdot3^{2i+1}\alpha \ (\mbox{mod}\ 2r) \mid 0\leq i < \frac{n-1}{2}\right\}\cap\left\{4\cdot3^{2i}b^{t-1}\alpha \ (\mbox{mod}\ 2r) \mid \frac{s}{2} \leq i \leq \frac{n-1}{2}\right\}=\emptyset.$$
We can choose
\begin{eqnarray*}
T_{d}&=&\bigcup_{a\in \Lambda_{d}}\Bigg(\bigg(\bigcup_{j=0}^{\frac{t-2}{2}}\left\{ 3^{2i+1}b^{2j}\alpha \ (\mbox{mod}\ 2r) \mid 0 \leq i <\frac{n-1}{2} \right\}\bigg)\\
&&\cup\bigg(\bigcup_{j=0}^{\frac{t-4}{2}}\left\{ 3^{2i}b^{2j+1}\alpha \ (\mbox{mod}\ 2r) \mid 0 \leq i <\frac{n-1}{2}\right\}\bigg)\\
&&\cup\left\{ 3^{2i}b^{t-1}\alpha \ (\mbox{mod}\ 2r) \mid \frac{s}{2} \leq i \leq \frac{n-1}{2}\right\}\Bigg)
\end{eqnarray*}
and $M_{4}^{'}(2d)\geq|T_{d}|=|\Lambda_{d}|\cdot\frac{ (n-1)(t-1)+n-s+1}{2} .$
\end{itemize}
\hfill$\blacksquare$\\

 For prime $r<100$, we give all the examples by Theorems 6,7 and 8 in Table 2.
\begin{table}[!h]
\caption{$M_{4}(2p)$ with $2\not\in\langle3\rangle_{p}$}
\centering
\begin{tabular}{|c|c|c|c|c|c|c|c|}
\hline
\hline
 $p$ & $n$ & $t$ & $s$ & $|\Lambda_{d}|$ & $M_{4}(2p)$ & $T_{d}$ & Thm. \\
 \hline
11 & 5 & 2 & 1 & 1 & $\geq3$ & $\{3,5,13\}$ & 8-(2) \\
13 & 3 & 4 & 2 & 1 & $\geq4$ & $\{3,7,15,25\}$ & 8-(2) \\
37 & 18 & 2 & 11 & 1 & $\geq16$ & $\{1, 5, 7, 9, 11, 17, 29, 31, 43, 45, 57, 63, 65, 67, 69, 73\}$ & 7-(2) \\
41 & 8 & 5 & 2 & 1 & $\geq16$ & $\{1, 5, 9, 11, 13, 17, 35, 37, 45, 47, 65, 69, 71, 73, 77, 81\}$ & 7-(1) \\
59 & 29 & 2 & 22 & 1 & $\geq18$ & $\{3, 5, 7, 11, 19, 25, 27, 29, 45, 51, 53, 63, 89, 93, 95, 99, 105, 107\}$ & 8-(2) \\
61 & 10 & 6 & 9 & 1 & $\geq25$ & \tabincell{c}{$\left\{1, 9, 13, 15, 19, 21, 23, 33, 43, 49, 53, 59, 65,\right.$ \\
$\left.  67, 75, 77, 81, 83, 85, 95, 97, 111, 115, 117, 119\right\}$} & 7-(2) \\
67 & 22 & 3 & 5 & 1 & $33$ & \tabincell{c}{$\left\{1, 9, 15, 17, 19, 21, 23, 25, 29, 33, 35, 37, 39, 47, 49, 55, 59, 65,\right.$ \\
$\left.  71, 73, 77, 81, 83, 89, 91, 93, 103, 107, 121, 123, 127, 129, 131\right\}$} & 6 \\
73 & 12 & 3 & 10 & 2 & $\geq24$ & \tabincell{c}{$\left\{1, 5, 9, 19, 21, 25, 33, 43, 45, 51, 65, 67, 79, 81,\right.$\\
$\left. 95, 101, 103, 113, 121, 125, 127, 137, 141, 145\right\}$} & 7-(1) \\
83 & 41 & 2 & 33 & 1 & $\geq37$ & \tabincell{c}{$\left\{3, 5, 17, 19, 21, 23, 25, 27, 29, 31, 33, 37, 39, 41,\right.$\\
45, 47, 49, 59, 67, 73, 77, 79, 85, 91, 95, 97, 101,\\
 $\left. 103, 105, 109, 113, 115, 131, 151, 153, 155, 159\right\}$} & 8-(2)\\
\hline
\end{tabular}
\end{table}

Since $M_{4}(2r)=\sum_{d\mid r}M_{4}^{'}(2d)$, the lower bounds on $M_{4}^{'}(2d)$ can imply a lower bound for $M_{4}(2r)$. Numerical result indicate that all the lower bounds on $M_{4}(2r)$ deduced from Theorems 3-8 are tight for $r\leq53$ and $\gcd(r,6)=1.$

$\mathbf{Conjecture:}$\ The lower bound of $M_{4}(2r)$ deduced from Theorems 3-8 are tight for all $\gcd(r,6)=1.$

\section{Summary}

In this paper, we are mainly consider the constructions of a maximal size $B_{1}[4](2^{k}r)$ set with $\gcd(r,6)=1.$ It can be applied to error correction for single asymmetric error of limited magnitude since all the syndromes are distinct. For $k\geq3$, we first reduce the construction of a maximal size $B_{1}[4](2^{k}r)$ set to a maximal size $B_{1}[4](2^{k-3}r)$ set which implies that we only need to determine a maximal size $B_{1}[4](4r)$, $B_{1}[4](2r)$ and $B_{1}[4](r)$ set.  The construction of a maximal size $B_{1}[4](4r)$ set has been completely solved in Theorem 2.  Furthermore, we discuss  maximal size $B_{1}[4](2r)$ set and given the calculation formula or lower bound of $M_{4}^{'}(2r)$. On the other hand, for $q\leq106$ we can determine all maximal size $B_{1}[4](q)$ sets by computer search. In all these examples, $M_{4}^{'}(2r)$ is equal to the lower bound in our result. It is reasonable to  conjecture that all the lower bounds are tight. We invite the readers to attack these open problems.



\begin{thebibliography}{1}
\vskip2mm
\bibitem{AAK} R. Ahlswede, H.  Aydinian and L.H. Khachatrian, ``Unidirectional errors control codes and related combinatorial problems,'' in \emph{Proc. 8th Int. Workshop on Algebraic and Comb. Coding Theory}, Tsarskoe Selo, Russia, Sep. 8-14, 2002, pp. 6--9.
\bibitem{AAKT1} R. Ahlswede, H.  Aydinian, L.H. Khachatrian and L.M.G.M Tolhuizen, ``On q-ary codes correcting allunidirectional errors of a limited magnitude,'' in \emph{Proc. 9th Int. Workshop on Algebraic and Comb. Coding Theory}, Kranevo, Bulgaria, Jun. 19-25, 2004, pp. 20--26.
\bibitem{AAKT2} R. Ahlswede, H.  Aydinian, L.H. Khachatrian and L.M.G.M Tolhuizen, ``On q-ary Codes Correcting All Unidirectional Errors of a Limited Magnitude,'' Jul. 2006. [Online]. Available: arXiv:cs/0607132v1 [cs.IT].
\bibitem{BE} S. Buzaglo and T. Etzion,``Tilings with n-dimensional chairs and their applications to asymmetric codes,'' \emph{IEEE Trans. Inf. Theory}, vol. 59,
no. 3, pp. 1573--1582, Mar. 2013.
\bibitem{CSB} Y. Cassuto, M. Schwartz, V. Bohossian and J. Bruck,. ``Codes for Asymmetric Limited-Magnitude Errors with Application to Multilevel Flash Memories,'' \emph{IEEE Trans. Inf. Theory}, vol. 56, no. 4, pp. 1582--1595, May 2010.
\bibitem{KBN}  T. Kl{\o}ve, B. Bose and N. Elarief, ``Systematic, single limited magnitude error correcting codes for Flash Memories,'' \emph{IEEE Trans. Inf. Theory}, vol. 57, no. 7, pp. 4477--4487, Aug. 2011.
\bibitem{KLNY} T. Kl{\o}ve, J. Luo, I. Naydenova, and S. Yari, ``Some codes correcting asymmetric errors of limited magnitude,'' \emph{IEEE Trans. Inf. Theory}, vol. 57, no. 11, pp. 7459--7472, Nov. 2011.
\bibitem{KLY} T. Kl{\o}ve, J. Luo and S. Yari, ``Codes correcting single errors of limited magnitude,'' \emph{IEEE Trans. Inf. Theory}, vol. 58, no. 4, pp. 2206--2219, Apr. 2012.
\bibitem{M}  M. Schwartz,``Quasi-cross lattice tilings with applications to flash
memory'' \emph{IEEE Trans. Inf. Theory}, vol. 58, no. 4, pp. 2397--2405, Apr. 2012.
\bibitem{XL} D. Xie and J. Luo, ``Optimal Equi-difference Conflict-avoiding Codes,'' Sep. 2018. [Online]. Available: arXiv.org: cs/1809.09300 [cs. IT].
\bibitem{YKB} S. Yari, T. Kl{\o}ve, and B. Bose,``Some codes correcting unbalanced errors of limited magnitude for flash memories'' \emph{IEEE Trans. Inf.
Theory}, vol. 59, no. 11, pp. 7278--7287, Nov. 2013.
\bibitem{ZG} T. Zhang and G. Ge,``New results on codes correcting single error of limited magnitude for flash memory'' \emph{IEEE Trans. Inf. Theory}, vol. 62,
no. 8, pp. 4494--4500, Aug. 2016.
\bibitem{ZZG} T. Zhang, X. Zhang, G. Ge, ``Splitter Sets and $k$-Radius Sequences,'' \emph{IEEE Trans. Inf. Theory}, vol. 63, no. 12, pp. 7633--7645, Dec. 2017.
\end{thebibliography}
\end{document}